\documentclass[aps,prd,preprint,floats,superscriptaddress]{revtex4}
\usepackage[reqno]{amsmath}
\usepackage{mathrsfs}
\usepackage{amssymb}

\usepackage[hypertex]{hyperref}
\usepackage{graphicx}

\newcommand{\be}{\begin{equation}}
\newcommand{\ee}{\end{equation}}
\newcommand{\bea}{\begin{eqnarray}}
\newcommand{\eea}{\end{eqnarray}}

\begin{document}

\pagestyle{plain}

\title{\begin{flushright} \texttt{UMD-PP-09-047}
\end{flushright}
Radiative Transmission of Lepton Flavor Hierarchies}
\author{Adisorn Adulpravitchai$^1$,~Manfred Lindner$^1$,\\
Alexander Merle$^1$,~and~Rabindra N.~Mohapatra$^2$
\\
{\normalsize \it $^1$Max-Planck-Institut f\"ur Kernphysik,}\\
{\normalsize \it Postfach 10 39 80, 69029 Heidelberg, Germany}\\
{\normalsize \it $^2$Maryland Center for Fundamental Physics and Department of}\\
{\normalsize \it Physics, University of Maryland, College Park, MD, 20742}\\
}
\date{\today}

\begin{abstract}
We discuss a one loop model for neutrino masses which leads to a
seesaw-like formula with the difference that the charged lepton masses
replace the unknown Dirac mass matrix present in the usual seesaw
case. This is a considerable reduction of parameters in the
neutrino sector and predicts a strong hierarchical pattern in the
right handed neutrino mass matrix that is easily derived from a
$U(1)_H$ family symmetry. The model is based on the left-right
gauge group with an additional $Z_4$ discrete symmetry which gives
vanishing neutrino Dirac masses and finite Majorana masses arising
at the one loop level. Furthermore, it is one of the few models that naturally allow for large (but not necessarily maximal) mixing angles in the lepton sector. A generalization of the model to the quark
sector requires three iso-spin singlet vector-like down type
quarks, as in $E_6$. The model predicts an inert doublet type scalar
dark matter.
 \end{abstract}

 \maketitle


\section{Introduction}
One of the major puzzles in particle physics beyond the standard
model (SM) is to understand the origin of neutrino
masses~\cite{review}. A simple paradigm is the seesaw
mechanism~\cite{seesaw} which introduces three right-handed (RH)
neutrinos with arbitrary Majorana masses additionally to the SM with the resulting
seesaw formula for the light neutrino mass matrix given by:
\begin{eqnarray}
{\cal M}_\nu~=~ -m_D^TM^{-1}_R m_D
\end{eqnarray}
The input values of $m_D$ and $M_R$ are then required to find the
neutrino masses. In the simple seesaw framework, the RH neutrino spectrum can therefore not be determined from neutrino
observations. Clearly, the knowledge of the right handed neutrino
spectrum would be of great phenomenological interest for testing the
model. If seesaw is embedded into grand unified theories it is
sometimes possible to predict $m_D$, so that one could get some
idea about the right handed neutrino masses. In this paper, we
present a bottom-up one loop scheme where we obtain the
following seesaw-like formula from a left-right symmetric model even though the Dirac mass matrix
vanishes to all orders in perturbation theory:
\begin{eqnarray}
{\cal
M}_\nu~=~\frac{\lambda'}{16\pi^2}M^{diag}_\ell M^{-1}_NM^{diag}_\ell
\end{eqnarray}
where $M^{diag}_\ell$ is the diagonal charged lepton mass matrix: $
M^{diag}_\ell={\rm diag}(m_e, m_\mu, m_\tau)$ and $\lambda'$ is a Higgs self
coupling.
As a result, the flavor structure of the RH neutrino mass matrix is
completely determined. We find a stronger hierarchy in the RH
neutrino sector compared to the charged leptons. Thus the radiative
corrections transmit the
charged lepton mass hierarchy into the RH neutrino sector (\emph{radiative transmission of hierarchies}). Furthermore the
hierarchy in the RH sector is such that it is easily obtainable from a
simple $U(1)_H$ family assignment. This is the main result of the paper.
As an application, we predict $B(\mu\to e+\gamma)$ in this model.

We also discuss how the quark sector can be made realistic since
the $Z_4$ symmetry leads to vanishing down quark
masses at tree level. Two ways to generate
realistic down quark masses and CKM angles are: (i) introduction
of color triplet iso-spin singlet fields that
give radiative masses to down quarks or (ii) the addition of three
iso-spin singlet vector-like down quarks which generate a tree
level mass for the down quarks. We only present the second scenario here,
which also has the property that it
leads to an inert doublet type scalar dark matter.

\section{The model}
Our model is based on the left-right (LR) symmetric group~\cite{LR}
$SU(2)_L\times SU(2)_R\times U(1)_{B-L}$ supplemented by a
discrete symmetry group $Z_4$. The quarks and leptons are assigned
as in the minimal LR model to left-right symmetric doublets. The
symmetry breaking is implemented also as in the minimal LR model by the
Higgs fields $\phi(2,2,0)$ and $\Delta_R (1,3,+2)\oplus
\Delta_L(3,1,+2)$.

In the leptonic sector of
the model, the $SU(2)_R\times U(1)_{B-L}$ breaking by the right
handed triplet with $B-L =2$ gives large Majorana masses to the RH
neutrinos~\cite{goran}. Unlike in the usual implementation of the seesaw formula
however, in our model, the Dirac mass for neutrinos vanishes to
all orders in perturbation theory due to the $Z_4$ symmetry, whose
effect on the various fields is given in the table below:
\begin{center}
\begin{tabular}{|c||c|}\hline Fields & $Z_4$ charge\\ \hline
$Q_R$ & $-i$\\ $L_R$ & $+i$ \\ $\phi$ & $+i$\\ $\tilde{\phi}\equiv
\tau_2\phi^*\tau_2$ & $-i$\\ $\Delta_R$ & $-1$ \\ \hline
\end{tabular}
 \end{center}
All other fields are assumed to be singlets of $Z_4$. The most
general potential for the left-right model has been discussed in the
literature before~\cite{desh}. The presence of the $Z_4$ symmetry in our
model forbids terms linear in the invariant ${\rm
Tr}(\tilde{\phi}^\dagger\phi)$ in the
potential
so that the minimum energy configuration corresponds to the following vev
for the $\phi$ field (instead of the general one in~\cite{LR}):
\begin{eqnarray}
\langle \phi\rangle~=~\left(\begin{array}{cc} \kappa & 0 \\ 0 & 0
\end{array}\right).
\end{eqnarray}
For the $\Delta_{L,R}$ fields we have:
\begin{eqnarray}
\langle\Delta^0_R\rangle~=~\begin{pmatrix}0 & 0 \\ v_R & 0\end{pmatrix},\;
\langle\Delta^0_L\rangle~=~0.
\end{eqnarray}
The gauge invariant Yukawa couplings of
the above $Z_4$ supplemented LR model are
\begin{equation}
{\cal L}_Y~=~h_q \bar{Q}_L\phi Q_R~+~h_l\bar{L}_L\tilde{\phi}L_R~+~\left[f(L_R^T \Delta_R L_R+L_L^T \Delta_L L_L)~+~h.c.\right].
\end{equation}
By an appropriate choice of the basis, we can choose both $h_{q,l}$ to be
diagonal matrices. There is no loss of generality in this. It is easy to
see that with the above assignment, we
get the Dirac neutrino mass $m_D=0$ and the diagonal Yukawa coupling
matrix $h_l$ is given by $h_l~=~{\rm diag}(m_e, m_\mu, m_\tau)/v_{wk}$. We also
note that there is no type II seesaw~\cite{type2} contribution to the
neutrino masses unlike in usual left-right models.

The stability of the minimum of the potential under radiative corrections can be seen as
follows: If we write $\phi \equiv (H, \eta)$ where $H$ and $\eta$
are two $SU(2)$ Higgs doublets with $Y=\mp 1$, then the above vev pattern
corresponds to $\langle H\rangle=v_{wk}\neq 0$ and $\langle \eta\rangle=0$. In
the language of the $\eta$ and $H$ fields, it is easy to see that the
Lagrangian of the model respects a remnant $Z_2$ symmetry under which
$\eta\to -\eta$ and $N_R\to -N_R$ and all other fields are singlets, so
that the zero vev for $\eta$ is protected by this symmetry.
Below the $SU(2)_R\times U(1)_{B-L}$ breaking scale, the model is a two
Higgs extension of the standard model that allows for small neutrino masses, similar to one
discussed in ref.~\cite{ma}.

\section{Seesaw-like formula for Neutrino masses}

\begin{figure}[t]
\centering
\begin{tabular}{lr}
 \includegraphics[width=8cm]{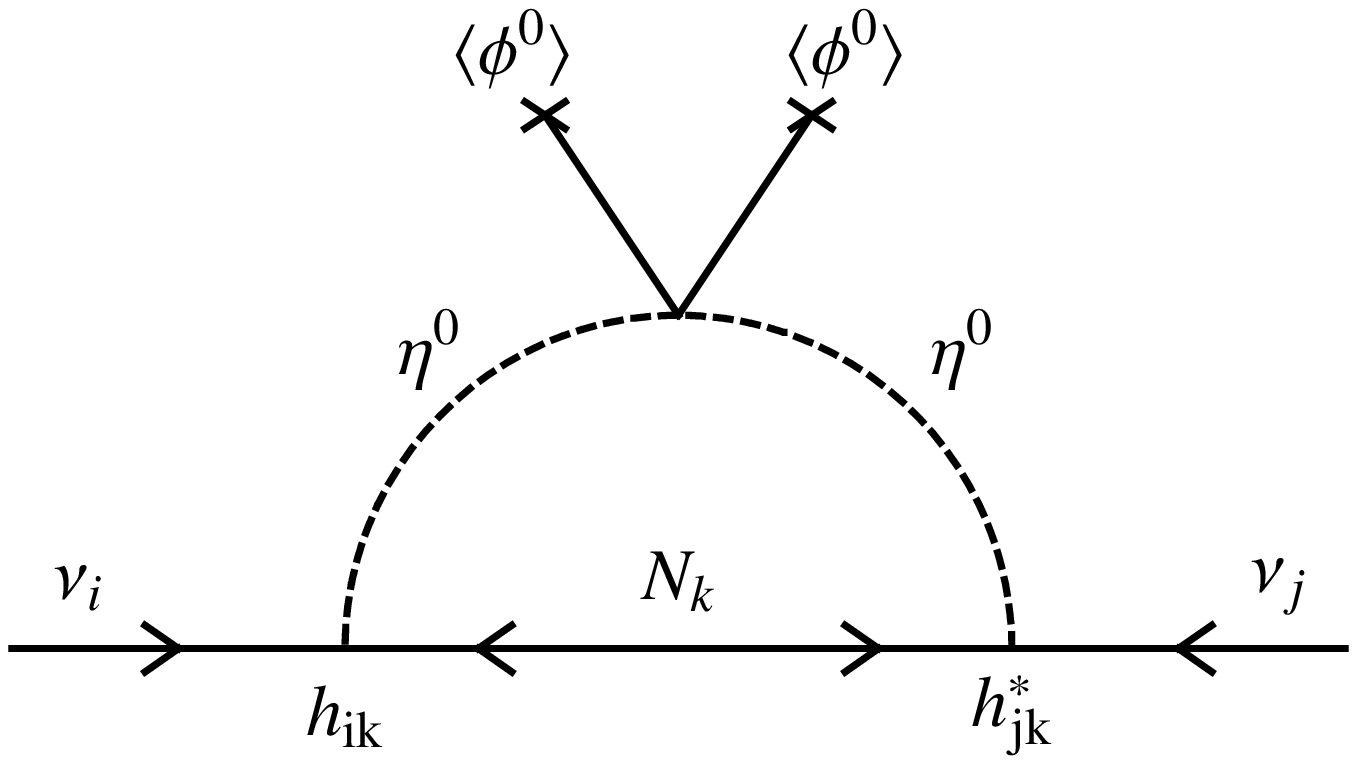} &
\includegraphics[width=8cm]{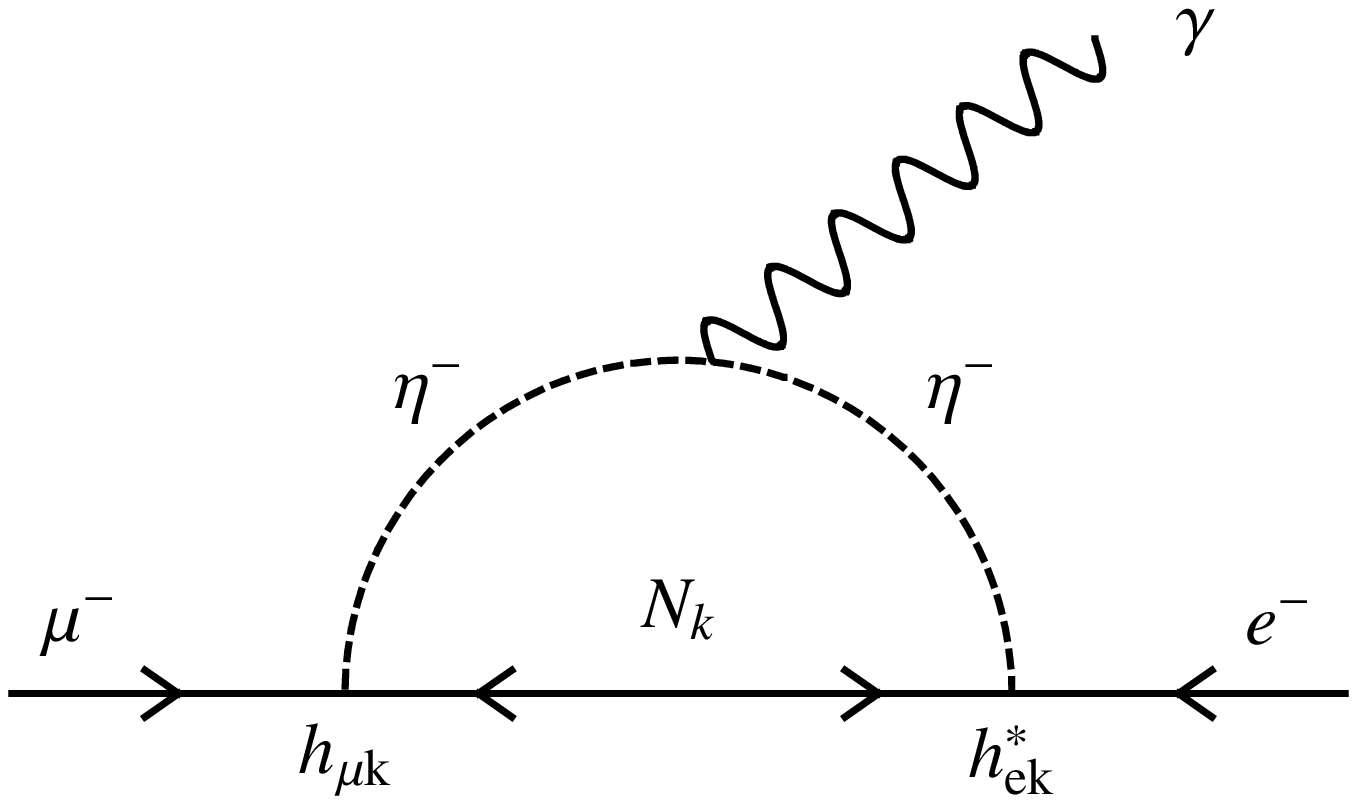}
\end{tabular}
 \caption{\label{fig:numass_LFV} The diagrams responsible for neutrino
 masses and for the rare decay $\mu \to e\gamma$ in a Ma-like model, cf.\
ref.~\cite{ma}.}
\end{figure}

As noted, at tree level, both
neutrino Dirac masses and the down
quark masses vanish. We will address the question of down quark masses
in the next section. As far as neutrinos are concerned, at one
loop level they pick up mass from the left diagram in
Fig.~\ref{fig:numass_LFV} with the neutrino
mass matrix given by the one loop formula
\begin{eqnarray}
{\cal M}_{\nu, ij} ~=~
\frac{1}{16\pi^2}m_{l,i}\Lambda_{ij}(\lambda',M_{N,ij})m_{l,j},
 \label{eq:M_nu}
\end{eqnarray}
where $\Lambda_{ij}$ is given by
\begin{equation}
 \frac{M_{N,ij}}{16 \pi^2} \left[ \frac{m^2(\sqrt{2} \Re
\eta^0)}{m^2(\sqrt{2} \Re \eta^0)-M_{N,ij}^2}
\log \left( \frac{m^2(\sqrt{2} \Re \eta^0)}{M_{N,ij}^2} \right)-
\frac{m^2(\sqrt{2} \Im \eta^0)}{m^2(\sqrt{2}  \Im \eta^0)-M_{N,ij}^2} \log
\left( \frac{m^2(\sqrt{2} \Im \eta^0)}{M_{N,ij}^2} \right) \right].
\end{equation}
The Higgs masses are given by
\begin{eqnarray}
 &&m^2(\sqrt{2} \Re \phi^0)=2\lambda_1 \kappa^2, \ M^2(\sqrt{2} \Re
\eta^0)=M_2^2+(\lambda_3+\lambda_4+\lambda_5)\kappa^2,\nonumber \\
 &&m^2(\sqrt{2} \Im \eta^0)=M_2^2+(\lambda_3+
\lambda_4-\lambda_5)\kappa^2,\ {\rm and}\ m^2(\eta^\pm)= M_2^2+\lambda_3
\kappa^2. \end{eqnarray}
Note that these couplings $\lambda_i$ are the effective couplings which
we get at low energies when the left-right symmetry is broken.

We assume that $m^2(\sqrt{2} \Re \eta^0) \ll M_{N,ij}^2$,
 $\Lambda_{ij}(\lambda',M_{N,ij}) \simeq 2 \frac{\lambda'}{M_{N,ij}^2}
 \log \left(\frac{M_{N,ij}^2}{m^2(\sqrt{2} \Re \eta^0)}\right)$, where $\lambda'$ is
equivalent to $\lambda_5$ in the Ma-model~\cite{ma}. Then, the light neutrino mass matrix can then be written as
\begin{eqnarray}
{\cal M}_{\nu, ij} ~=~
 \frac{2 \lambda'}{16\pi^2}m_{l,i}\left(M^{-1}_N\right)_{ij}m_{l,j}
\log(M_{N,ij}^2/m^2(\sqrt{2} \Re \eta^0)).
\end{eqnarray}
Note that we can absorb $\log \left(\frac{M_{N,ij}^2}{m^2(\sqrt{2}  \Re
\eta^0)}\right)$ into $\left(M^{-1}_N\right)_{ij}$ without loss of generality.

Since we have a rough idea about the form of the neutrino mass matrix in
the limit of zero CP phase and small reactor angle $\theta_{13}$, we can
use it to get an idea about the elements of the RH neutrino mass matrix.
It is interesting that all elements of this mass matrix can be determined.

\subsubsection{Normal Hierarchy}
The neutrino mixing observables~\cite{neutrinomixing} we use are:
\begin{eqnarray}
& & \Delta m_{21}^2=7.65 \times 10^{-5} \mbox{~eV}^2,\
|\Delta m_{31}^2|=2.40 \times 10^{-3} \mbox{~eV}^2 ,\ \nonumber \\
& & \sin^2 \theta_{12}=0.304,\ \sin^2 \theta_{23}=0.5, \ {\rm and}\
\sin^2 \theta_{13}=0.01.
\end{eqnarray}
The charged lepton masses we take from ref.~\cite{QuarkFit}:
\begin{equation}
m_e=0.511, \; \; m_{\mu}=105.658, \; \; \mbox{~and~} m_{\tau}=1776.84 \;
\;\; \mbox{MeV}.
\end{equation}

To fit the neutrino oscillation data, we can use
\begin{equation}
M_N = \frac{2 \lambda'}{16\pi^2}  \begin{pmatrix}
 1.83 \times 10^{6} & -1.76 \times 10^{8} & 2.87 \times 10^{9} \\
 \times & 1.80 \times 10^{10} & -2.91 \times 10^{11}\\
 \times & \times & 4.81 \times 10^{12}
 \end{pmatrix}  \mbox{~GeV} ,
 \label{eq:MN_NH}
\end{equation}
where the mass eigenvalues are given by $(M_{N1},M_{N2},M_{N3})=\frac{2 \lambda'}{16\pi^2} (9.55 \times
 10^{4},4.65 \times 10^{8},4.83 \times 10^{12})$ GeV. Note that, in order to
avoid the $N$ detection in the $Z$-boson decay width, $\lambda'$ has to be larger than $0.0037$.

The neutrino masses are given by
\begin{equation}
m_1=0.0001, \; \; m_2=0.0087, \; \mbox{~and~} m_3=0.049 \;\;
\mbox{eV}.
\end{equation}

\subsubsection{Inverted Hierarchy}
To fit the neutrino oscillation data, we can use
\begin{equation}
M_N = \frac{2 \lambda'}{16\pi^2}  \begin{pmatrix}
 5.57 \times 10^{3} & -3.80 \times 10^{6} & 6.42 \times 10^{7}\\
 \times & 5.59 \times 10^{10} & 9.37 \times 10^{11}\\
 \times & \times & 1.58 \times 10^{13}
 \end{pmatrix} \mbox{~GeV} ,
 \label{eq:MN_IH}
\end{equation}
where the mass eigenvalues are given by $(M_{N1},M_{N2},M_{N3})=\frac{2 \lambda'}{16\pi^2} (5.31 \times 10^{3},4.48 \times 10^{8},1.59 \times 10^{13})$ GeV, where $\lambda'$ now has to be larger than $0.67$.

The neutrino masses are
\begin{equation}
m_1=0.049, \; \; m_2=0.050, \; \mbox{~and~} m_3=0.0001 \;\; \mbox{eV}.
\end{equation}

Note that in both cases, there is a strong hierarchy in the RH neutrino
sector in a way similar to the charged lepton sector. This is what we
label as the \emph{radiative transmission of hierarchy} from charged leptons to
the RH neutrinos. Note that this mechanism, given a certain form of $M_N$ (with small mixings), naturally allows for large mixing angles in the SM lepton sector, that are not necessarily maximal. This is different from many other models, where in most cases only zero or maximal mixing is predicted. Note however, that there are also exceptions to this: E.g., the size of the mixing angle could be determined by underlying discrete symmetries~\cite{Blum:2007nt}, or it could arise from an anarchical pattern of the neutrino mass matrix~\cite{Haba:2000be}.

To see analytically why this happens,
let us try to reconstruct $M_N$ from the tri-bimaximal form for the
PMNS-matrix~\cite{Goswami:2009yy},
\begin{equation}
 U_{\rm PMNS}=
 \begin{pmatrix}
 \sqrt{\frac{2}{3}} & \frac{1}{\sqrt{3}} & 0 \\
 -\frac{1}{\sqrt{6}} & \frac{1}{\sqrt{3}} & -\frac{1}{\sqrt{2}} \\
 -\frac{1}{\sqrt{6}} & \frac{1}{\sqrt{3}} & \frac{1}{\sqrt{2}}
 \end{pmatrix}.
 \label{eq:UPMNS}
\end{equation}
Using this and Eq.~\eqref{eq:M_nu}, we can write down $M_N$ as function
of $\lambda'$ and of the light neutrino mass eigenvalues $m_{1,2,3}$. It
is given by $\frac{\lambda'}{6 m_1 m_2 m_3}$ times
\begin{equation}
 \begin{pmatrix}
 2 (m_1+2m_2) m_3 m_e^2     & 2 (m_1-m_2) m_3 m_e m_\mu & 2 (m_1-m_2) m_3 m_e m_\tau \\
 2 (m_1-m_2) m_3 m_e m_\mu  & (3 m_1 m_2+m_2 m_3+ 2 m_1 m_3 ) m_\mu^2 & (-3 m_1 m_2+m_2 m_3+ 2 m_1 m_3 ) m_\mu m_\tau \\
 2 (m_1-m_2) m_3 m_e m_\tau & (-3 m_1 m_2+m_2 m_3+ 2 m_1 m_3 ) m_\mu m_\mu & (3 m_1 m_2+m_2 m_3+ 2 m_1 m_3 ) m_\tau^2
 \end{pmatrix}.
 \label{eq:MN}
\end{equation}
 If we assume normal ($m_1=p^2 m_0$, $m_2=p m_0$, and $m_3=m_0$, with
 small $p$) or inverted hierarchy ($m_1=m_0$, $m_2=m_0$, and $m_3=p
m_0$), the corresponding matrices will roughly look like
\begin{equation}
 (M_N)_{\rm NH}=\frac{\lambda'}{6 p^2 m_0}
 \begin{pmatrix}
 4 m_e^2      & -2 m_e m_\mu    & -2 m_e m_\tau \\
 -2 m_e m_\mu  & m_\mu^2      & m_\mu m_\tau \\
 -2 m_e m_\tau & m_\mu m_\tau & m_\tau^2
 \end{pmatrix}
 \label{eq:MN_NH_form}
\end{equation}
and
\begin{equation}
 (M_N)_{\rm IH}=\frac{\lambda'}{2 p m_0}
 \begin{pmatrix}
 2p m_e^2      & 0   & 0 \\
 0  & m_\mu^2      & -m_\mu m_\tau \\
 0 & - m_\mu m_\tau & m_\tau^2
 \end{pmatrix}.
 \label{eq:MN_IH_form}
\end{equation}
 Note that the reconstruction of all matrices
 (Eqs.~\eqref{eq:MN_NH_form},
 and~\eqref{eq:MN_IH_form}) has led us to heavy neutrino mass matrices
 which are hierarchical and stiff. In all cases, having a light neutrino
mass close to zero ($p\rightarrow 0$ in Eqs.~\eqref{eq:MN_NH_form}
 and~\eqref{eq:MN_IH_form}) can only increase this hierarchy, but not
 destroy it. Especially in Eq.~\eqref{eq:MN_IH_form} the 11-entry is
 fixed, which means that we will generically have one fixed RH neutrino
 mass that is not too heavy. A similar situation happens for the
quasi-degenerate case.

These mass matrices for RH neutrinos have a structure
that is easily obtainable from
the Froggat-Nielsen (FN) mechanism~\cite{Froggatt:1978nt} with a
$U(1)_H$ family symmetry
with $H$ charges $(0,1,2)$ for the third, second, and the first
generation right handed lepton doublets. The left-right and $U(1)_H$
invariant
Yukawa couplings in this case can be written as:
\begin{eqnarray}
{\cal
L}_{Y,H}~=~h^l_3\bar{L}_{3,L}\tilde{\phi}L_{3,R}+
h^l_2\bar{L}_{2,L}\tilde{\phi}L_{2,R}\frac{\sigma}{M}+
h^l_1\bar{L}_{1,L}\tilde{\phi}L_{1,R}\left(\frac{\sigma}{M}\right)^2\\
\nonumber
+\left[ \sum_{a,b=1,2,3}
f_{ab}{L}^T_{a,R}\tilde{\Delta}L_{b,R}
\left(\frac{\sigma}{M}\right)^{6-(a+b)} + h.c.\right].
\end{eqnarray}
For an appropriate choice of $\frac{<\sigma>}{M}$ (roughly $1/20$
in the normal hierarchy case), we get the desired hierarchy in
both the charged lepton masses as well as in the RH neutrino
sector. This hierarchy then translates into a structure of the
light neutrino mass matrix that naturally yields large mixing
angles, although no values are excluded a priori.

We can also give a prediction for $\mu \to e
\gamma$~\cite{Adulpravitchai:2009gi}, which is transmitted by
the heavy neutrinos (cf.\ right diagram of Fig.~\ref{fig:numass_LFV}): The Yukawa coupling in the basis
where the heavy neutrino mass matrix is diagonal is given by $h=U^{-1}
{\rm diag}(m_e,m_\mu,m_\tau)/v_{wk}$, where $U$ is the matrix that diagonalizes $M_N$. For a charged Higgs mass of 100~GeV and
$\lambda'=0.7$, the prediction for ${\rm Br}(\mu \to e \gamma)$ is
$6\cdot 10^{-16}$ for normal and $8\cdot 10^{-16}$ for inverted ordering,
where we have used Eqs.~\eqref{eq:MN_NH} and~\eqref{eq:MN_IH}. If we go
to  smaller values for $\lambda'$, the branching ratio increases
($3\cdot 10^{-12}$ for $\lambda'=0.01$ and normal ordering), which might
be very interesting in light of the upcoming MEG experiment~\cite{Ritt:2006cg}.

\section{Extension to Quark sector}
It is clear from Eq.~(4) that at the tree level in our model, only
the up quarks are massive. We present two ways to make the quark
sector realistic by giving mass to the down quarks, (i) one where the
$Z_2$ symmetry, that keeps Dirac mass of the neutrino to be zero, is
softly broken and (ii) another one by adding three vector-like down
quarks, where we can keep the $Z_2$ symmetry exact. We only discuss the
second option here.

For (ii), we extend the model by adding three $SU(2)_{L,R}$ singlet, color triplet,
$B-L=2/3$ quarks (denoted by $D_{L,R}$) and two Higgs doublets under
the $SU(2)_{L,R}$ groups with $B-L=1$ (denoted by $\chi_{L,R}$). Under the
$Z_4$ symmetry, the $\chi_{L,R}$ and $D_R$ are invariant, whereas $D_L\to
-iD_L$. It is easy to write down a potential for $\chi_{L,R}$ with
asymmetric mass terms for them so that they have
nonzero VEVs. Since the discrete symmetry does not allow the term
$\chi^\dagger_L\phi\chi_R$ term in the potential, the additional fields do not
destabilize the $\phi$ vev pattern assumed in the bulk of the paper. The new
Yukawa interaction
that is invariant under $Z_4$ and gauge symmetry is given by
\begin{eqnarray}
{\cal L}_{new}~=~f_D(\bar{Q}_L\chi_LD_R+\bar{Q}_R\chi_RD_L) + h.c.
\end{eqnarray}
After spontaneous symmetry breaking the down quarks now have masses
where they pair with the new down quarks (rather than the usual ones of the
SM). As a result, the $SU(2)_R$ partner of the up quark is a heavy down
quark unlike in the minimal left-right model~\cite{LR}. In fact, after
symmetry breaking, one could write the left and right doublets as follows:
$Q_L = (u_L,d_L)$ and $Q_R = (u_R, D_R)$ ($D_R$ and $d_R$ swap roles), where the mass of $D$ is in the
10 to 100 TeV range. We emphasize that there is no direct mass term
between $D_L$ and $D_R$.

To fit the down quark masses and the CKM matrix, the Yukawa coupling need to be
\begin{equation}
f_D=\begin{pmatrix}
 0.89 & 24.7 & 14.1\\
 \times & 106.5 & 169.9\\
 \times & \times & 4192.9
 \end{pmatrix} \frac{1}{v_L},
\end{equation}
where $v_L$ is the vev of $\chi_L$. This appears to be a completely viable way to generate down quark
masses. An interesting feature of this model is that the surviving $Z_2$
remains an exact symmetry, and
as result the neutral member of the second doublet in $\phi$ can act
as dark matter~\cite{Dolle:2009fn}, since it couples to quarks as $\phi^0_2
\bar{d}_LD_R$, and as long as $m_{\phi^0_2}\ll M_D$, the $\phi^0_2$ is
 stable with stability guaranteed by the $Z_2$
symmetry~\cite{inert}.

\section{Conclusion}
In summary, we have shown that a radiative one loop model for neutrino
masses proposed in~\cite{ma} arises as a low energy limit of a left-right
model which then provides a natural explanation of the two elements of
the~\cite{ma} proposal: (a) the reason for the extra doublet with its
particular discrete symmetry property and (b) the origin of the right
handed neutrino mass. Furthermore, the radiative transmission of hierarchies makes large but non-maximal mixing angles in the leptonic sector plausible. Left-right embedding also reduces the number of
parameters in the model, making it predictive in the hadronic and leptonic flavor
sectors.

The work of R.~N.~M. is supported by the US National Science Foundation
under grant No. PHY-0652363 and Alexander von Humboldt Award (2005 Senior Humboldt
Award). One of the authors (R.~N.~M.) is grateful to Manfred Lindner for hospitality at the Max-Planck-Institut f\"ur Kernphysik in Heidelberg during the time when part of the work was
was carried out. This work has been supported by the DFG-Sonderforschungsbereich Transregio 27 ``Neutrinos and beyond -- Weakly interacting particles in Physics, Astrophysics and Cosmology''.

\end{document}